\documentstyle[12pt,psfig]{article}
\begin{document}

\baselineskip=.22in
\renewcommand{\baselinestretch}{1.2}
\renewcommand{\theequation}{\thesection.\arabic{equation}}

\begin{flushright}
{\tt hep-th/0208140}
\end{flushright}

\vspace{5mm}

\begin{center}
{\Large \bf  Vacuum Structure and Global Strings\\[1mm]
with Conical Singularities}\\[12mm]
{Yoonbai Kim\footnote{yoonbai$@$skku.ac.kr}, 
O-Kab Kwon\footnote{okwon$@$newton.skku.ac.kr}, 
Jongsu Sohn\footnote{jongsu$@$newton.skku.ac.kr}}\\[3mm]
{\it BK21 Physics Research Division and Institute of Basic Science,\\
Sungkyunkwan University, Suwon 440-746, Korea}
\end{center}

\vspace{5mm}

\begin{abstract}
Vacuum structure and global cosmic strings are analyzed
in the effective theory of  self-interacting O(2) scalar fields
on (3+1)-manifolds with conical singularities. In the context of one-loop
effective action computed by heat-kernel methods with $\zeta$-function 
regularization, we find an inhomogeneous vacuum of minimum energy and
suggest some reason why low-energy global strings are likely to be 
generated at the conical singularities. 
\end{abstract}


\newpage

\setcounter{equation}{0}
\section{\large\bf Introduction}
Cosmic strings remain somewhere in our universe as relics of cosmological
phase transition at early times. These linear topological defects have been 
extensively studied in connection with cosmological problems~\cite{Vil}.
Global strings were proposed as a candidate of producing primordial density
perturbations leading to galaxy formation~\cite{VE}. However, the original
purpose seems unlikely to be achieved due to radiation of Goldstone bosons 
around oscillating closed loops of strings~\cite{Dav} and other stringent
astrophysical bounds~\cite{LH}.
Recently other physical applications are also investigated, {\it e.g.}, time 
evolution of global string network~\cite{YY} and possible generation of 
several types of global strings in high density QCD~\cite{FZ}.

When infinite straight U(1) global strings are coupled to Einstein gravity, 
static solutions with cylindrical symmetry encounter unavoidable 
singularity~\cite{Gre}. With a negative cosmological constant, 
global vortices become
seeds for charged BTZ black holes in (2+1) dimensions without 
physical singularity~\cite{KKK}. Such
spacetime structure with horizons is extended to a viable model of
6-dimensional Randall-Sundrum type brane worlds, which is composed of 
our world on the 3-brane and 2 extra-dimensions identified with spatial 
plane of global vortices~\cite{CK}.
 
The questions we are interested in in this paper are the vacuum structure 
and the global strings in the effective theory with 
conical singularities. Suppose that an infinite straight local cosmic string
was generated throughout a cosmological phase transition in the very early 
universe, {\it e.g.}, a Nielsen-Olesen vortex string at grand unification 
(GUT) scale. If there has been no other phase transition up to the late-time
transition of a very low energy, {\it e.g.}, electroweak (EW) scale or lower 
scale, the supermassive local cosmic string may be observed through
some gravitational effects, i.e., background spacetime has a deficit angle.
Even though a phase transition occurs at the low energy, released latent heat
or other fluctuations may be not enough to {\it melt} this high-energy 
cosmic string. So the survived string may affect the new vacuum structure of 
spontaneous symmetry breaking. In section 2, we assume an O(2) scalar theory 
with conical singularity in the background spacetime for tractability, and
compute one-loop effective action including a $\delta$-function singularity.
In section 3, we analyze vacuum of the effective theory and show that an
inhomogeneous vacuum, jumping from symmetric local maximum at the
points of conical singularities to broken global minimum at the other space,
is minimum energy configuration
instead of homogeneous broken vacuum. In section 4, we address our main issue
that the production of a global string along the supermassive local cosmic
string seems favorable and natural.

The obtained low-energy global string containing the supermassive
local cosmic string core looks like an infinite straight candle 
with a heavy wick.
This kind of possibilities was firstly proposed in the case of 
superconducting strings, {\it e.g.},
the O(10) string with an inner core of $\tilde{\rm U}
(1)$ magnetic flux~\cite{Wit}. 
It may also be understood as a simplified field-theoretic calculation of
daily life experience that the structure of extend objects can easily 
formed at the site of a
defect or dust when a vapor (or a liquid) is cooled down to the liquid
(or the solid). 

\section{\large\bf Effective action in the presence of a cosmic string}
Suppose that a straight cosmic string was produced throughout a 
cosmological
phase transition in the very early universe. When its 
species is a local cosmic string, {\it e.g.}, a string-like object of 
a Nielsen-Olesen vortex of Abelian Higgs model, the energy density is localized
around the stringy core of the cosmic string, of which size is the 
inverse of the Higgs mass. 
At an energy scale much lower than that of the Higgs mass,  
the cosmic string looks like an extremely-thin straight wire so that its 
energy-momentum tensor can be approximated as a $\delta$-function on a plane
orthogonal to the string direction~:
\begin{equation}\label{emt}
T^{\mu}_{\;\;\nu}\sim M_{0}^{2} {\rm diag}(1,0,0,1)\delta(x)\delta(y),
\end{equation}
where string direction is chosen by $z$-axis and mass of the string per 
unit length, $M_{0}^{2}$, is roughly of order the square of the scale of 
the phase transition.

The existence of such massive object in the early universe is likely to be
detected by gravitational effect. Spacetime structure of the cosmic string
(\ref{emt}) is determined by Einstein equations and then, in cylindrical 
coordinates, metric is expressed as follows
\begin{equation}\label{conmetric}
ds^{2}=-dt^{2}+dr^{2}+r^{2}(1-4GM_{0}^{2})^{2}d\theta^{2}+dz^{2}.
\end{equation}
This spacetime is flat everywhere except for the site of the string core at
$r=0$, and the effect of the local cosmic string is detected only by a 
deficit angle 
$2\pi-\gamma =8\pi GM_{0}^2$ on the $(x,y)$-plane (or $(r,\theta)$-plane)
$C_{\gamma}$. Except for supermassive scale ($4GM_{0}^{2}\ge 1$), $C_{\gamma}$ 
describes nothing but a cone~\cite{Got}.
 
{}From now on let us consider a model at low energy in the presence of a
very massive straight cosmic string. To be specific, we take into
account massless O(2) linear sigma model in the gravitational background 
of a straight local cosmic string~(\ref{conmetric}), described by the 
action 
\begin{equation}\label{act1}
S= \int_{{\cal M}} d^{4}x \left(
-\frac12 \partial_\mu\phi_a\partial^\mu\phi_a
- \frac{\lambda}{24} (\phi^2)^2 \right),
\end{equation}
where $\phi^2 = \phi_a\phi_a \;(a=1,2)$ and the base manifold is ${\cal M}
=M^{2}\times C_{\gamma}$. 
The situation with which we are dealing should involve two ingredients : One
is a phase transition and the other some effect of high-energy cosmic
string skeleton. In this paper, we will compute effective action $\Gamma_{\rm
tot}(\phi)$ so that the vacuum shifts from symmetric vacuum $\langle \phi
\rangle =0$ to new broken vacuum $\langle \phi\rangle \ne 0$ and a few 
singular terms are induced in effective potential, due to $\delta$-function 
singularity at the string site. 
It is actually enough to calculate the one-loop effective potential 
in order to fulfill the above demands up to leading order.

Here we compute the one-loop effective action $\Gamma^{(1)}$ by using 
background field methods near classical field $\phi_a^c$.
In the context of Euclidean path integral formalism, the one-loop contribution
$\Gamma^{(1)}$ to the effective action is 
\begin{eqnarray}
\Gamma^{(1)}(\phi_c) &=& -\ln \int {\cal D}\phi^q \; 
e^{-\frac12 \int_{{\cal M}_{\rm E}} d^4 x\; \phi_a^q A_{ab} \phi_b^q}
\label{effact1}\\
&=& \frac12\; \ln\det\left(\frac{A}{\mu^2}\right).
\label{effaction2}
\end{eqnarray} 
In the second line of Eq.~(\ref{effaction2}) an arbitrary mass parameter 
$\mu$ was introduced to keep the effective action dimensionless.
Physicswisely, it will play a role of an ultraviolet cutoff and will be
identified as the scale of the phase transition.
The specific form of the operator $A$ after Euclideanization is
\begin{equation}\label{ope}
A_{ab}=-\left(\partial_t^2 +\partial_z^2 +\Delta_c - 
\frac16\lambda (\phi^c)^2\right)\delta_{ab} 
+ \frac\lambda 3 \phi^c_a\phi^c_b ,
\end{equation}
where under the metric (\ref{conmetric}) the Laplace-Beltrami operator 
$\Delta_c$ in Eq.~(\ref{ope}) is expressed by 
\begin{equation}
\Delta_c = \partial^2_r + \frac1r \partial_r + \frac1{r^2} 
\partial^2_\theta.
\end{equation} 

It is well-known that $\zeta$-function regularization is an effective scheme
for the calculation of the one-loop effective action 
(\ref{effaction2})~\cite{BD}.
In this method the $\zeta$-function with eigenvalues $\lambda_j$'s 
of the differential operator $A$ takes the form
\begin{eqnarray}
\zeta(s|A)&=& \sum_j \lambda_j^{-s} = \frac1{\Gamma(s)} 
\sum_j \int^\infty_0 d\tau\;\tau^{s-1} e^{-\lambda_j\tau} 
\nonumber \\
&=& \frac1{\Gamma(s)}\int^\infty_0 d\tau\;\tau^{s-1}\int_{{\cal M}_{\rm E}}
 d^4x\;K(x,x,\tau),
\label{zeta}
\end{eqnarray} 
where the kernel $K(x,y,\tau)\equiv \langle x|e^{-\tau A}|y \rangle $ 
satisfies the equation
\begin{equation}
-\partial_\tau K(x,y,\tau) = A K(x,y,\tau),
\end{equation}
and its boundary condition is given by a $\delta$-function 
$\delta^4_{{\cal M}_{\rm E}}(x-y)$
on the manifold ${\cal M}_{\rm E}=R^{2}\times C_{\gamma}$ 
\begin{equation}\label{deltafun}
\lim_{\tau\to 0} K(x,y,\tau) = \delta^4_{{\cal M}_{\rm E}}(x-y).
\end{equation}
Therefore the one-loop effective action (\ref{effaction2}) is 
expressed in terms of the $\zeta$-function (\ref{zeta})
\begin{equation}\label{effaction3}
\Gamma^{(1)}(\phi_c) = -\frac12 \zeta'(0|A) 
-\frac12 \ln \mu^2 \zeta(0|A),
\end{equation}
where $\zeta'(0|A) = \frac {d\zeta(s|A)}{ds}|_{s=0}$.

Though the background spacetime is spatially inhomogeneous and static vortex 
configurations will be taken into account at the subsequent sections, we 
compute the one-loop effective potential in this section. 
This calculation is enough 
for the above requirement such as the phase transition and the effective 
terms due to conical singularities. 
Now let us turn off spacetime dependence of 
the fields $\phi_{a}$, and denote the classical field $\phi_c$ as $\phi$
in what follows for simplicity.
Then the $\zeta$-function in the one-loop effective action (\ref{effaction3})
is simplified as follows
\begin{eqnarray}
\zeta(s|A) &=& \frac1{\Gamma(s)} \int^\infty_0 d\tau\; 
\tau^{s-1} \int_{{\cal M}_{\rm E}} d^4 x\; {\rm tr} \langle x| 
e^{-\tau\left[-(\partial^2_t +\partial^2_z +\Delta_c 
- \frac\lambda 6 \phi^2)\delta_{ab} 
+ \frac\lambda 3 \phi_a\phi_b
\right]}|x \rangle\nonumber \\
&=& \frac{V_2}{4\pi \Gamma(s)} \int^\infty_0 d\tau\; \tau^{s-2}
\int_{C_\gamma} d^2 x\;\langle x| e^{\tau \Delta_c} 
|x\rangle 
\left(e^{-\frac{\tau\lambda}6 \phi^2} + e^{-\frac{\tau\lambda}2 \phi^2} 
\right) , 
\label{zetafun}
\end{eqnarray}
where `tr' denotes the trace over internal index $a,b$, and $V_2$ is 
2-dimensional volume of the $(t,z)$-coordinates.
Using a well-known result for the case with conical singularities~\cite{Dow}
\begin{equation}\label{zetacon}
\int_{C_\gamma} d^2 x\; \langle x| e^{-\tau(-\Delta_c)}|x\rangle = 
\frac{V(C_\gamma)}{4\pi\tau} + \frac1{12} 
\left( \frac{2\pi}{\gamma} - \frac{\gamma}{2\pi}\right),  
\end{equation}
we have 
\begin{eqnarray}\label{zetafun2}
\zeta(s|A) &=&\frac{V_2 V(C_\gamma)}{16\pi^2 (s-1)(s-2)} 
\left[\left(\frac{\lambda\phi^2}6\right)^{2-s} 
+\left(\frac{\lambda\phi^2}2\right)^{2-s}\right]\nonumber \\ 
&&+ \frac{V_2 \gamma_0}{16\pi^2(s-1)}\left[\left(\frac{\lambda\phi^2}6
\right)^{1-s} + \left(\frac{\lambda\phi^2}2 \right)^{1-s}\right],
\end{eqnarray}
where $V(C_\gamma)$ is volume of the cone $C_\gamma$ 
and $\gamma_0 = \frac{\pi}3 
\left( \frac{2\pi}{\gamma} - \frac{\gamma}{2\pi}\right)$.
Therefore, substitution of Eq.~(\ref{zetafun2}) into Eq.~(\ref{effaction3})
leads to the one-loop effective potential
\begin{equation}\label{effaction4}
\Gamma^{(1)}(\phi) = \frac{5V_2 V(C_\gamma)}{1152 \pi^2} 
\lambda^2 \phi^4 \left[\ln \left(\frac{\lambda \phi^2}{2 \mu^2}\right) 
-\frac1{10} \ln 3 -\frac32\right] + \frac{V_2 \gamma_0}{48\pi^2} 
\lambda \phi^2 \left[1 + \frac14 \ln 3 
-\ln \left(\frac{\lambda \phi^2}{2 \mu^2}\right)\right].
\end{equation}

We obtain an effective action by adding
the classical action (\ref{act1}) and the one-loop effective 
potential (\ref{effaction4}), which coincides with that containing the leading 
one-loop quantum correction in the derivative expansion
\begin{eqnarray}
\Gamma_{{\rm tot}}(\phi) = \int_{{\cal M}}d^4x \left\{ 
-\frac12 \partial_\mu \phi_a\partial^\mu\phi_a 
-V_{{\rm eff}}(\phi)
-\frac{ \gamma_0\delta_c(x)}{48\pi^2} 
\lambda \phi^2 \left[1 + \frac14 \ln 3   
-\ln \left(\frac{\lambda \phi^2}{2 \mu^2}\right)\right] \right\} ,
\label{totaction}
\end{eqnarray}
where $\delta_c(x)$ is $\delta$-function on the cone $C_\gamma$ and
$V_{{\rm eff}}(\phi)$ is regular part of the effective potential~:
\begin{equation}\label{pot}
V_{{\rm eff}}(\phi) = \frac\lambda{24} \phi^4 +
\frac{5\lambda^2}{1152 \pi^2}
\phi^4 \left[\ln \left(\frac{\lambda \phi^2}{2 \mu^2}\right)
-\frac1{10} \ln 3 -\frac32\right]
+ \frac{5 \mu^4}{192\cdot 3^{\frac45}\pi^2} \exp\left(2-
\frac{96\pi^2}{5\lambda}\right).
\end{equation}
In Eq.~(\ref{pot}) we adjusted a cosmological constant to zero at its true 
minimum and subtracted the energy of the cosmic string
since it does not change physics of our interest.

\section{\large\bf Inhomogeneous vacuum due to conical singularities}
As well-known, the effective potential $V_{\rm eff}(\phi)$ 
in Eq.~(\ref{pot}) comprises a local maximum at $\phi=0$
with a positive cosmological constant and the true vacuum is attained
at 
\begin{equation}\label{vacvalue}
\phi (\equiv v)= \sqrt{\frac{2\mu^2}{\lambda}}\; \exp\left[\frac{\lambda
(10+ \ln 3) - 96 \pi^2}{20\lambda}\right].
\end{equation}
An intriguing observation should be noted. 
At such broken vacuum 
with vanishing cosmological constant, the energy
is not zero even for the homogeneous vacuum configuration of $\langle
\phi \rangle=v$ but includes additional positive contribution from the site 
of the cosmic string $(r=0)$, which is proportional to the length of the string.
The energy per unit length along $z$-axis for the broken vacuum is
\begin{equation}\label{sten}
\left. E_{z}\right|_{\phi=v}
= \frac{\gamma_0 ( 192\pi^2 +3\lambda \ln 3)}
{160\cdot 3^{\frac9{10}} \pi^2 } 
\exp\left(1-\frac{48\pi^2}{5\lambda}\right)\frac{\mu^{2}}{\lambda}>0.
\end{equation}

To minimize its
energy, the vacuum configuration should be static and 
cylindrically-symmetric along the $z$-axis.
Let us look into another possibility of inhomogeneous vacuum that 
the scalar field $\phi$ has $r$-dependence. 
Introducing dimensionless quantities
\begin{eqnarray}\label{anv}
&&\phi(x)=\frac{\mu}{\sqrt{\lambda}}f(r)~~~
\left(\tilde{v} =\sqrt{2}\; \exp\left[\frac{\lambda
(10+ \ln 3) - 96 \pi^2}{20\lambda}\right]\right), \\
\label{var}
&&r \rightarrow \tilde{r} = \mu r, ~~~
\delta_c(x) \rightarrow \tilde{\delta}(\tilde{x})
= \frac1{\mu^2} \delta_c(x), 
\end{eqnarray}
we express the energy per unit length along the $z$-axis such as
\begin{eqnarray}
E_z = \frac{\mu^2}\lambda\int^\infty_0d\tilde{r}\;
\tilde{r}\int^\gamma_0 d\theta
\left[ \frac12 \left(\frac{d f}{d \tilde{r}}\right)^2
+ \tilde{V}_{{\rm eff}}(f)
+\tilde{\delta}(\tilde{x})\; U(f) \right],
\label{vacen}
\end{eqnarray}
where dimensionless part of the effective action (\ref{pot}) becomes
\begin{eqnarray}\label{pot2}
\tilde{V}_{{\rm eff}}(f) &=&
\frac{f^4}{24} +\frac{5\lambda}{1152\pi^2} f^4
\left(\ln \frac{f^2}2-\frac1{10} \ln 3 -\frac32\right)
+\frac{5 \lambda}{192\cdot 3^{\frac45}\pi^2} \exp \left[2\left(1-
\frac{48\pi^2}{5\lambda}\right)\right], \nonumber \\ \\
U(f) &=& \frac {\gamma_0\lambda}{48 \pi^2}  f^2
\left(1+\frac14 \ln 3 - \ln \frac {f^2}2 \right).\label{dpot}
\end{eqnarray} 
{}From here on, we denote the dimensionless quantities without tilde notation 
for convenience. 

Outside the core of the cosmic string $(r>0)$, the singular part of the 
effective potential, $\delta(r) U(f)/\gamma r$, does not contribute to the 
energy functional (\ref{vacen}). Consider the one-parameter family of 
configurations,
\begin{equation}\label{sca}
f_\kappa (r) \equiv f(\kappa r).
\end{equation}
If $f(r)$ be the vacuum at $r>0$, then $f(r)$ is an extremum of the 
energy functional. 
Since the derivative term in Eq.~(\ref{vacen}) is invariant under the 
scaling, the extremum 
condition provides $V_{{\rm eff}}(f) =0$ so that $f=v$ should be the 
vacuum at $r>0$.
As mentioned previously, extension of the vacuum $f=v$ to the origin 
has positive energy contribution from the singular potential 
like as (\ref{sten}).
The inhomogeneous vacuum would be given by a stationary solution 
of equation of motion
\begin{equation}\label{vaceqn}
\frac1r \frac d{dr} r \frac{df}{dr}
-\frac{dV_{{\rm eff}}(f)}{df}
=\frac{\delta(r)}{r} \frac{d U(f)}{df}. 
\end{equation}
Since second term at left-hand side of Eq.~(\ref{vaceqn}) is
finite at the origin, integration of it from zero to an infinitesimal 
$\varepsilon$ gives
\begin{equation}\label{vaceqn2}
\varepsilon \left. \frac{df(r)}{dr}\right|_{r=\varepsilon} 
= \left. \frac{dU(f(r))}{df}\right|_{r=0}.
\end{equation}
Right-hand side of Eq.~(\ref{vaceqn2}) is a finite constant so that its 
solutions are classified into two~:
\begin{equation}\label{vsol}
\lim_{\varepsilon\rightarrow 0}f(\varepsilon)
\left\{
\begin{array}{ll}
\sim\ln\varepsilon & \mbox{when~}\left.\frac{dU}{df}\right|_{r=0}\ne 0\\
=0~{\rm or}~\sqrt[8]{48} &\mbox{when~}\left.\frac{dU}{df}\right|_{r=0}=0
\end{array}
\right.
.
\end{equation}
For the singular solution proportional to $\ln\varepsilon$, the energy 
per unit length along the $z$-axis (\ref{vacen}) is  logarithmically divergent  
due to the derivative term in Eq.~(\ref{vacen}), i.e.,
$E_{z}\sim\lim_{\varepsilon\rightarrow 0}\int^{\varepsilon}dr\, r
(d\ln r/dr)^{2}+\cdots \sim \ln\varepsilon $~\cite{Jac}.
So this cannot be a vacuum solution. 
Note that $f(0) = \sqrt[8]{48}$ is also ruled out because 
the valid range of $f$ 
should not be much larger than $v$. 
Therefore, the remaining possibility in Eq.~(\ref{vsol}) 
is $f(0)=0$.\footnote{Though 
$f \rightarrow \infty$ looks like global minimum 
with negative infinite energy in Eq.~(\ref{vacen}), 
$f \gg \mu/\sqrt{\lambda}$ is unphysical because the
valid range of $f$ in the computation of effective potential 
(\ref{effaction4}) does
not allow such limit.}
Since the $\delta$-function term is also invariant under 
the scale transformation (\ref{sca}), the scaling behavior of the 
energy functional (\ref{vacen}) tells us that the energetically favored 
configuration is 
\begin{equation}\label{fvac}
f(r) = 
\left\{
\begin{array}{ll}
0 & \mbox{at~} r= 0\\
v & \mbox{at~} r\ne 0
\end{array}
\right.
.
\end{equation}
For this configuration, both the regular and singular terms of the 
effective potential in Eq.~(\ref{vacen}) have vanishing energy contribution.
One may easily notice that the derivative term also does not contribute 
to the energy according to the following estimation
\begin{equation}
\int^\varepsilon_0 dr\; r\; \frac12 \left( \frac{df}{dr}\right)^2 \sim 
\left(\varepsilon\left.\frac{df}{dr}\right|_{r=\varepsilon}\right)^2 = 0.
\end{equation}

Now we find the true vacuum of minimum energy. 
When the phase transition was accomplished so that the broken vacuum is 
realized almost everywhere, the site of the supermassive cosmic string 
still remains in the symmetric phase. 
Amazingly enough, such nonanalytic inhomogeneous vacuum has zero energy. 
Suppose that the core region of the symmetric phase swells 
and becomes connected 
smoothly to the broken vacuum at asymptotic region. 
Then, the radiation of massless Goldstone bosons is naturally 
initiated~\cite{Dav} and its final destination is expected to be the 
obtained nonanalytic vacuum (\ref{fvac}). 

\section{\large\bf Global string with massive cosmic string wick}
In the previous section we have shown that the
effective action obtained in section 2 favors the  
symmetry-broken vacuum $\langle \phi\rangle=v$ almost everywhere but does
the symmetric vacuum $\langle\phi\rangle =0$ at the site of 
the massive cosmic string wick. Since a representative topological defect of
this O(2) scalar theory is global vortex string, let us study infinite
global vortex-line along  $z$-axis, which connects smoothly the symmetric
vacuum at the vortex center and the broken vacuum at exterior region of
its core. 

Field configuration of the global string is static
and cylindrically-symmetric along the $z$-axis in order to minimize 
its energy. For an appropriate 
ansatz to make the field single-valued 
\begin{equation}\label{fldansz}
\phi_1 + i \phi_2 = \frac\mu{\sqrt{\lambda}}\;e^{i\bar{n} \theta}f(r),
\end{equation}
the energy per unit length along the $z$-axis involves a topological term 
proportional to $\bar{n}^2$~;
\begin{eqnarray}
E_z 
= \frac{\mu^2}\lambda\gamma\int^\infty_0d\tilde{r}\;
\tilde{r}
\left[ \frac12 \left(\frac{d f}{d \tilde{r}}\right)^2
+\frac{\bar{n}^2}{2\tilde{r}^2}f^2 
+ \tilde{V}_{{\rm eff}}(f)
+\frac {\gamma_0 }{48 \pi^2\gamma}\frac{\delta(\tilde{x})}
{\tilde{r}} \lambda f^2
\left(1+\frac14 \ln 3 - \ln \frac {f^2}2 \right) \right],\nonumber\\
\label{energy}
\end{eqnarray}
where $\bar{n} =2\pi n/\gamma$ and $n$ is an integer.
Though effect of the deficit angle appears in the ansatz of the global 
vortex (\ref{fldansz}), topological charge $\nu$ of it is an 
integer $n$, i.e., $\nu \equiv \oint_{\partial C_\gamma } 
d\vec{l}\cdot (\phi_1\vec{\partial}\phi_2 - \phi_2\vec{\partial}\phi_1)/
2\pi \phi^2 = n$.

\begin{figure}\label{bacsfig}
\begin{center}{
\input{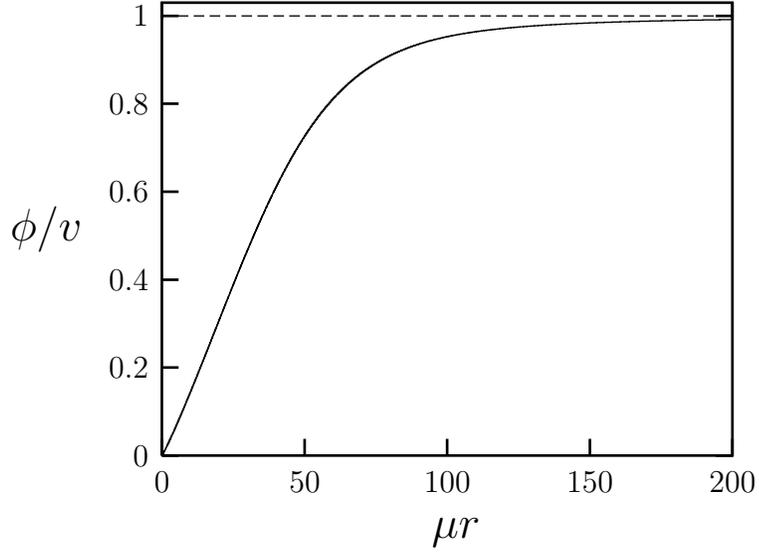}
}
\end{center}
\begin{center}{\large (a)}
\end{center}
\begin{center}{
\input{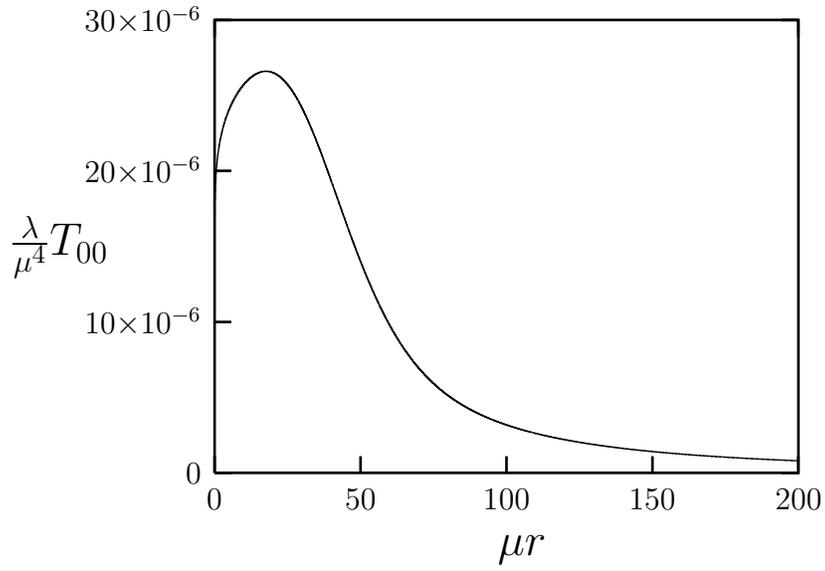}
}
\end{center}
\begin{center}{\large (b)}
\end{center}
\caption{Profile of a global vortex when
$\bar{n}=1.1$ and $\lambda=20$~: (a) scalar amplitude $\phi$, (b) energy 
density $T_{00}$.}
\end{figure}

To let the configuration nonsingular, we have a boundary 
conditions at the origin, $f(r=0) =0$,
and, to make energy density vanish at spatial infinity, 
the scalar amplitude should approach to the broken vacuum at asymptotic 
region, $f(r=\infty) = v$. 
Now our task is to find a vortex string configuration 
interpolating these two boundary conditions and satisfying 
the equation of motion
\begin{equation}\label{feqn1}
\frac{d^2 f}{d r^2}= - \frac 1r \frac{d f}{dr}
+ \frac{\bar{n}^2}{r^2} f -\frac d {df}(-V_{{\rm eff}}),
\end{equation}
where $V_{\rm eff}(f)$ is given in Eq.~(\ref{pot2}).
Series expansion of $f(r)$ near the origin is
\begin{eqnarray}\label{f0}
f(r) \simeq  f_0 r^{\bar{n}} 
+ \frac{5\lambda f_0^3\bar{n}}
{576\pi^2(\bar{n}+1)(2 \bar{n} +1)} r^{3\bar{n} + 2} \ln r + \cdots ,  
\end{eqnarray}
where $f_0$ is a constant determined by the proper behavior of the
fields at the asymptotic region. Though the configuration seems not to 
be regular at the origin due to the singularity at the vertex of the cone, 
Eq.~(\ref{f0}) tells us that there is no $\delta$-function-like energy 
addition at the origin at all. The behavior at large $r$ is obtained 
precisely by analyzing the equation of motion (\ref{feqn1}) 
\begin{eqnarray}\label{rinf}
f(r) \simeq v - 
\frac{144\pi^2\bar{n}^2}{ 5\lambda v r^2}
- \frac{10368 \pi^4 \bar{n}^2(3\bar{n}^2 + 8)}{25\lambda^2 
v^3 r^4}+ \cdots .
\end{eqnarray}
Since we cannot have an exact solution, a numerical solution is shown in 
Fig.~\ref{bacsfig}.\footnote{Existence and 
uniqueness of the static global vortex solution can easily be proved. Once 
the following identifications, $f\to x,\; r\to t$ and
$V_{{\rm eff}}(f) \to -V(x)$, are made, Eq.~(\ref{feqn1}) is analogous to 
a Newtonian equation of a unit-mass particle in 1-dimensional motion, of which
applied forces are time-dependent friction, time-dependent repulsion, and 
conservative force in order in Eq.~(\ref{feqn1}). 
The motion corresponding to the vortex solution is
that starting from the valley of the potential $(x=0)$ at $t=0$ 
and arriving at the hilltop $(x=v)$ after infinite time $t$ elapsed.} 
Profile of the energy density shows a long range tail in
addition to a ring shape at the vortex core as expected. 
Existence of such long range 
term can easily be read from the energy expression (\ref{energy})~: 
The second term proportional to $\bar{n}^{2}$ is ${\cal O}(1/r^{2})$
at large distance from the global cosmic string core so that the energy 
per unit length along the $z$-axis is logarithmically divergent as ordinary 
global strings did.

Now discussion on the stability of the above global string is in order
under some simplified situations.
Suppose that a phase transition occurred in the very early universe, 
{\it e.g.},
at the GUT scale ($10^{16}~$ GeV), and topologically-stable local cosmic 
strings have been produced via spontaneous symmetry breaking. If there does 
not happen subsequent phase transition up to the EW scale (1 TeV)
like desert in the standard model and we neglect complicated dynamics 
of cosmic strings~\cite{Vil}, the high-energy cosmic string wick 
may survive even throughout the EW phase transition because released 
latent heat may not be enough to melt the extremely-massive topological 
defects. Specifically, 1 TeV seems too small compared to $10^{16}$ GeV.
So form of the remnant near each high-energy cosmic string is likely 
to be observed as a line of the deficit angle of the background spacetime 
or Aharonov-Bohm effect of a  {\it magnetic} flux tube. 
This may justify the usage of
our calculation of the effective potential in the section 2.
Since the background planar space $C_{\gamma}$ orthogonal to the string 
direction has a circle as its spatial boundary 
$\partial C_\gamma = S^1$, the low-energy global strings 
we obtained are also topologically stable despite its 
logarithmically-divergent energy. 

An intriguing issue in this stage may be to answer the question whether
or not a straight global string favors its generation site along the cosmic
string wick. Here we deal with energy difference between two possible
configurations without considering the huge mass of the background string
wire, which is always same irrespective of existence of the low-energy strings
so that it does not contribute to the energy difference. Suppose that a
global string of unit topological charge ($n=1$) is produced along the
cosmic string wick. Its core size is roughly estimated from the first
and second terms of Eq.~(\ref{rinf}), i.e.,
$\sqrt{\lambda} v\, r_{\rm core}\sim (1+4GM^{2}_{0})12\pi/\sqrt{5}$.
The core energy per unit length along the $z$-axis is computed by an integration
of Eq.~(\ref{energy}) from $r=0$ to $r=r_{\rm core}$ after inserting $f(r)=0$
into the integrands : $E_{z}^{\rm core}\sim \frac{\mu^{2}}{\lambda}\gamma
\int_{0}^{r_{\rm core}}dr r \, V_{\rm eff}(f=0)=(1+4GM^{2}_{0})\pi v^{2}/16$.
The logarithmically divergent energy is obtained by another integration
of Eq.~(\ref{energy}) from $r=r_{\rm core}$ to an infra-red cutoff $R$
after substituting the vacuum value $f(r)=v$ : $E_{z}^{R}\sim
\frac{\mu^{2}}{\lambda}\gamma\int_{r_{\rm core}}^{R}dr r \,
(\bar{n}^{2}/2r^{2})|_{n=1}=(1+4GM^{2}_{0})\pi v^{2}\ln(R/r_{\rm core})$.
Therefore, both the core radius and the energy per unit length along the
wick are $(1+4GM^{2}_{0})$ times those generated outside the singularity.
Since $4GM_0^2 \sim 10^{-6}$ at the GUT scale, energy difference can be
negligible except for the cosmic strings of Planck scale, $4GM_0^2\sim 1$.
When a global string is generated, the boundary conditions of the strings
coincide with those of the vacuum (\ref{fvac}) at both the origin and 
spatial infinity.
So generation of a global string is nothing but a process that the
extremely-thin wire starts to swell naturally to the extended core of radius
$r_{\rm core}$ carrying a topological charge $n$. The energy supplied for such
process may be almost the same as the energy of a global string given above.
However, in order to produce a global string at the homogeneous broken vacuum,
the process is rather complicated~\cite{Vil}, some more energy cost may be 
needed to excite a defect in addition to that of a global string. 
In the context of symmetry,
the former preserves the cylindrical symmetry but the latter does not.
These suggest that an infinite straight global string
along the supermassive thin cosmic string wick is a favorable configuration,
which can naturally be generated throughout a low-energy phase transition.

\section{\large\bf Concluding remarks}
We have analyzed vacuum structure and global string solutions in the
one-loop effective action of self-interacting O(2) scalar fields of which
background space has conical singularities. In the one-loop effective 
potential, a singular part proportional to 2-dimensional $\delta$-function 
emerges in addition to well-known regular part of it. 
The configuration of minimum energy is an inhomogeneous vacuum in which the 
homogeneous broken vacuum dominates almost everywhere but the vertex points of 
conical singularities have symmetric vacuum. If the region of symmetric vacuum
swells and then is connected smoothly to broken vacuum due to a topological 
winding, it becomes a hybrid of global string of low energy and local 
cosmic string wick of high energy. 

Since our computation seems not sufficient for supporting all the steps of
the procedure of consideration, refined calculation is needed in future. 
What we obtained in this paper can be extended to other similar cases, 
{\it e.g.}, local strings of gauge theories with a local cosmic string wick,
global strings from dynamical symmetry breaking, monopoles with a supermassive
global monopole seed. The obtained topological defects of candle shape
with a wick should be applied for understanding various cosmological 
questions as the others have been examined.
 
\section*{\large\bf Acknowledgments}
We would like to thank Chanju Kim for valuable comments and discussions.
Y.K. wishes to acknowledge informative discussions with T. Vachaspati, 
H.-C. Kim, and B. K. Lee at the early stage of this work.
This work was the result of research activities (Astrophysical Research
Center for the Structure and Evolution of the Cosmos (ARCSEC))
supported by Korea Science $\&$ Engineering Foundation.
O-K. Kwon is supported by the KOSEF fellowship.

\end{document}